# Faint Extended X-ray sources: A new Large Area Survey

L.R. Jones[1,5], C.A. Scharf[1,5], H. Ebeling[2]

E. Perlman[1], M. Malkan[3], G. Wegner[4]

[1] *Code 668, NASA/Goddard Space Flight Center, MD 20771.*
[2] *Institue of Astronomy, Madingley Rd, Cambridge, UK.*
[3] *Dept. of Astronomy, UCLA, Los Angeles, CA 90024.*
[4] *Dept. Physics & Astronomy, Dartmouth College, 6127 Wilder Lab, Hanover, NH 03755.*

**Abstract.** We have searched for extended X-ray sources in the ROSAT PSPC archive using the Voronoi Tesslation and Percolation algorithm (VTP). This algorithm detects all significant sources regardless of shape or size, and is particularly suited to finding low surface brightness extended sources which could be missed or wrongly interpreted by other algorithms. In an initial 13 sq deg (66 fields) we have found 22 sources with detected flux $>7 \times 10^{-14}$ erg cm$^{-2}$ s$^{-1}$ (0.5-2 keV) and sizes of 1 to 5 arcmin. Optically, they range from a single bright nearby galaxy which has been resolved, an Abell cluster which is revealed to have two (probably merging) components, and groups and clusters of galaxies at estimated redshifts beyond z=0.4. This work is part of an ongoing survey with the aims of (a) measuring the low luminosity ($<10^{44}$ erg s$^{-1}$), high redshift (z>0.2) X-ray luminosity function of clusters, and (b) investigating cluster morphologies and unusual systems (eg 'fossil' groups).

## 1. Introduction

Understanding the formation and evolution of clusters of galaxies is of fundamental importance to the study of the growth of structure in the Universe. Complete samples of clusters and groups with which to make such studies can be most reliably constructed using X-ray detection techniques, avoiding the projection effects and selection effects inherent in optical surveys. Cold Dark Matter and other hierarchical models predict that the rate of growth of structure in the Universe was such that massive clusters containing a hot intracluster medium should have formed at late times - $z \leq 0.3$. The two current surveys (Edge et al. 1990, Henry et al. 1992) broadly confirm this scenario, with evidence for fewer luminous clusters at z>0.2 than locally. However, the cluster X-ray luminosity function (XLF) at luminosities lower than $\approx 10^{44}$ erg s$^{-1}$ has not been measured at z>0.15 using a complete, X-ray selected sample. The WARPS (Wide Angle ROSAT Pointed Survey) cluster survey is designed to make this measurement, testing whether the high redshift XLF has a Schechter function shape like the low redshift XLF (Ebeling et al. 1995), and measuring the form of the XLF evolution. The WARPS survey includes both extended sources and point X-ray sources with galaxy counterparts, ensuring that no clusters are missed because of their small size or large redshift. Here we concentrate on the X-ray extended sources.

---
[5]NRC Research Associate

## 2. The VTP Algorithm

The VTP algorithm is a general method for the detection of non-Poissonian structure in a distribution of points (Ebeling & Wiedenmann 1993). Briefly, Voronoi cells are constructed around each raw photon in the image. The reciprocal of the cell size corresponds to a measure of local surface brightness. The background level is determined from the distribution of cell areas compared to that expected from a Poissonian photon distribution. Significant photons in cells smaller than a threshold above the background are flagged and neighbouring photons collected into sources. The background is then recomputed and significant photons redetermined in an iterative procedure. *All* sources, point-like and extended, are found by VTP, with no binning or smoothing size required. We then use the second moments of the photons in each source to discriminate between (a) point-like sources and (b) both extended and blended sources. All extended and blended sources are checked by eye and compared to the results of a point source search in order to distinguish the true extended sources. As a test, VTP was run on fields containing the following known optical clusters. All the clusters listed were detected as extended sources:

Table 1. Known clusters detected as extended X-ray sources

| Cluster | ref | redshift | Note |
|---|---|---|---|
| Pavo | Griffiths et al 1992 | 0.13 | In Einstein deep survey field |
| J1888.16CL | Couch et al 1991 | 0.56 | SGP |
| J1836.10RC | Couch et al 1991 | 0.275 | |
| F1767.10TC | Couch et al 1991 | 0.664 | |

## 3. Results

We searched the inner 15 arcmin radius of 66 randomly selected ROSAT PSPC fields (13 sq deg) with exposures >10 ksec at $|b|$ >20°, and found 22 serendipitous extended sources with detected flux >7x10$^{-14}$ erg cm$^{-2}$ s$^{-1}$ (0.5-2 keV). This flux limit gave signal/noise values of 7-20. The size of the X-ray sources vary from 1 to 5 arcmin in their longest dimension. Simulations show that we are complete at this flux limit for a reasonable range of cluster sizes (core radius <500 kpc at z>0.15). The simulations also show that at our typical surface brightness limit, we detect >50% of the total flux of a cluster at the flux limit of the survey with a King profile of $r_c = 250$ kpc, $\beta = 0.66$ and redshift z>0.1.

We have checked the NED database, at the positions of these sources. Of the 22 extended sources, only 4 are already catalogued. One of these is Abell 2465, which is detected as two separate extended X-ray sources at the same redshift of z=0.245. The two components are separated by $1.5h^{-1}$ Mpc and may be in the process of merging. The other catalogued sources are a nearby resolved elliptical galaxy, an extended radio source, and a cluster of galaxies at z≈0.4. Many of the 18 uncatalogued extended X-ray sources are coincident with groups or clusters of galaxies on optical survey plates, including relatively bright groups of galaxies and other faint, probably distant, clusters (see fig 1). For those sources with no counterpart on the plates, or only a faint counterpart within 1 mag of the plate limit, we have obtained an R band CCD image at the Lick Observatory 1m Nickel telescope or the MDM 1.3m or 2.4m telescopes. Initial spectroscopy has also been performed at the Lick 3m telescope. In fig 1 we show some of these images.

## 4. Discussion

The ROSAT PSPC images have sufficient signal/noise that a much lower limiting surface brightness can be reached than in the EMSS (Henry et al 1992). Thus a better estimate of the total flux of sources at the flux limit of the EMSS is obtained, as well as the ability to probe to fainter limits. The VTP algorithm fully exploits this signal/noise in an automated source search. Algorithms based on the intrument psf would underestimate the source flux, or may miss the low surface brightness sources entirely.

The number of sources detected here is slightly higher than that predicted by the LogN-LogS relations in Rosati et al. (1995), although a complete analysis is still underway. The main conclusions of this paper are that (a) large numbers of clusters and groups of galaxies are detectable in the ROSAT PSPC archive and (b) VTP provides an excellent way of both finding them and measuring theeir fluxes accurately.

Further optical follow-ups and extensions to more ROSAT fields and fainter fluxes are in progress.

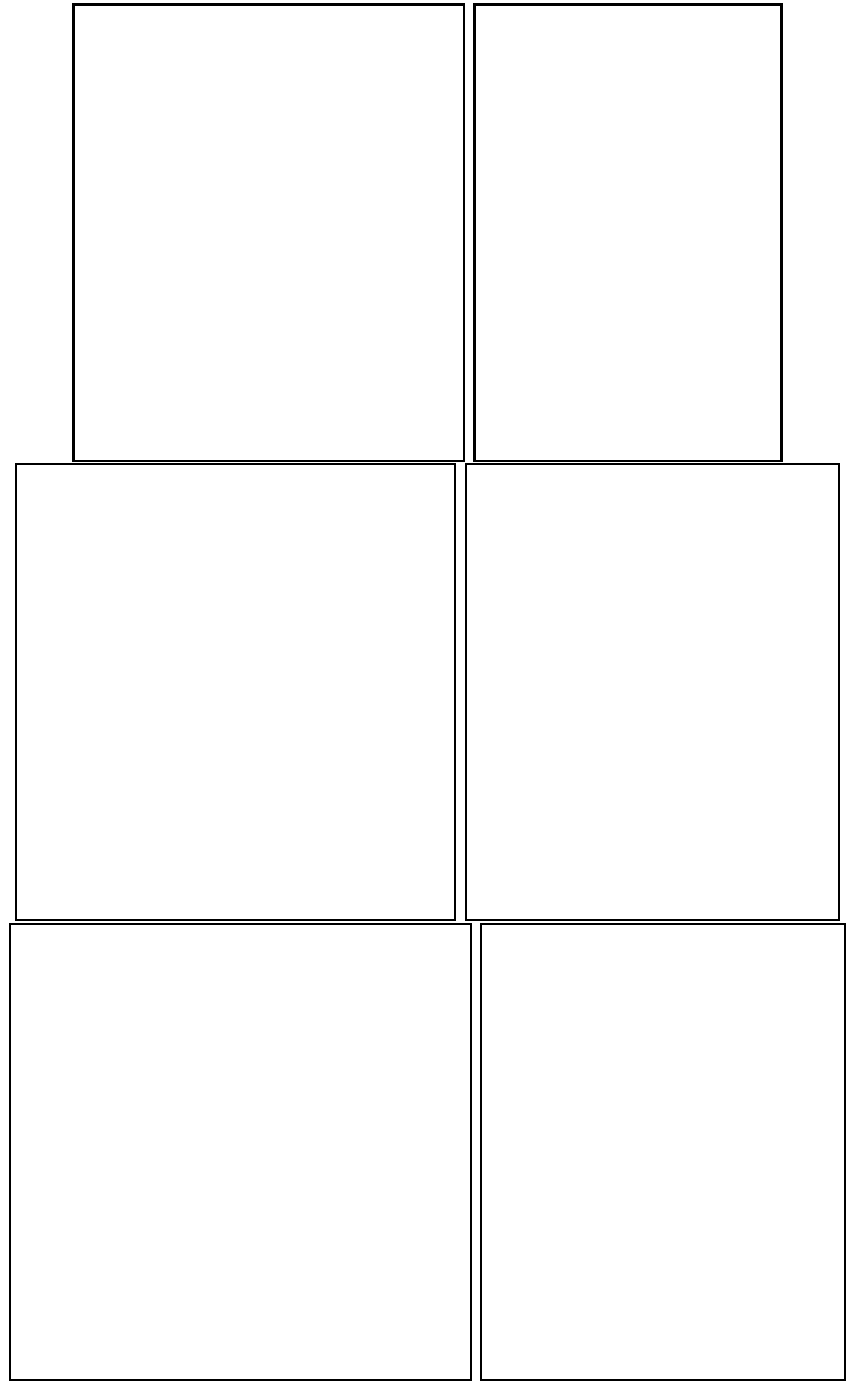

Figure 1. R band images of the counterparts of examples of extended X-ray sources. The X-ray centroid is at the centre of each image, which is approximately 2 arcmin across. The two components of Abell 2465 are shown at the top. The limiting magnitude is R=21.5 mag.